\documentclass[conference]{IEEEtran}

\usepackage{cite}
\usepackage{amsmath,amssymb,amsfonts}
\usepackage{algorithmic}
\usepackage{graphicx}
\usepackage{textcomp}
\usepackage{lipsum} 
\usepackage{xcolor}
\def\BibTeX{{\rm B\kern-.05em{\sc i\kern-.025em b}\kern-.08em
    T\kern-.1667em\lower.7ex\hbox{E}\kern-.125emX}}
\usepackage{float}
\usepackage{caption}
\usepackage{subcaption}
\usepackage[nolist]{acronym}
\usepackage{diagbox} 
\usepackage{enumitem}
\usepackage[none]{hyphenat}

\newcommand{\iu}{\textrm{j}}
\newcommand{\vect}[1]{\boldsymbol{\mathbf{#1}}}

\def\Ttran{\mathsf{T}}
\def\imagunit{\mathsf{j}} 

\begin{document}

\title{An Enhanced Polar-Domain Dictionary Design for Elevated BSs in Near-Field U-MIMO\vspace{-0.3cm}}

\author{
\IEEEauthorblockN{L. Antonelli, A. A. D'Amico, L. Sanguinetti}
\IEEEauthorblockA{Department of Information Engineering, University of Pisa, Pisa, Italy \\
National Inter-University Consortium for Telecommunications (CNIT), Parma, Italy\vspace{-0.7cm}}
}

\maketitle

\begin{acronym}
    \acro{THz}{TeraHertz}
    \acro{GHz}{GigaHertz}
    \acro{MHz}{MegaHertz}
    \acro{UE}{user equipment}
    \acro{i.i.d.}{indipendent-and-identically-distributed}
    \acro{SNR}{Signal to Noise Ratio}
     \acro{NMSE}{Normalized Mean Square estimation
    Error}
    \acro{LoS}{line-of-sight}
    \acro{RF}{radio frequency}
    \acro{U-MIMO}{Ultra-massive MIMO}
    \acro{MIMO}{Multiple Output Multiple Input}
    \acro{BS}{base station}
    \acro{ULA}{uniform linear array}
    \acro{FC}{Fully-Connected}
    \acro{P-SOMP}{polar simultaneous orthogonal matching pursuit}
    \acro{FDP}{Fully Digital P-SOMP}
    \acro{P-SIGW}{Polar Simultaneous Gridless Weighted}
    \acro{SE}{spectral efficiency}
    \acro{CSI}{channel state information}
    \acro{RoI}{region-of-interest}
    \acro{DoF}{degree-of-freedom}
    \acro{GP}{ground plane}
    \acro{MR}{maximum ratio}
    \acro{MMSE}{minimum mean square error}
\end{acronym}

\acresetall

\begin{abstract}


Near-field \ac{U-MIMO} communications require carefully optimized sampling grids in both angular and distance domains. However, most existing grid design methods neglect the influence of base station height, assuming instead that the base station is positioned at ground level—a simplification that rarely reflects real-world deployments. To overcome this limitation, we propose a generalized grid design framework that accommodates arbitrary base station locations. Unlike conventional correlation-based approaches, our method optimizes the grid based on the minimization of the optimal normalized mean squared error, leading to more accurate channel representation. We evaluate the performance of a hybrid U-MIMO system operating at sub-THz frequencies, considering the \ac{P-SOMP} algorithm for channel estimation. Analytical and numerical results show that the proposed design enhances both channel estimation accuracy and spectral efficiency compared to existing alternatives.


\end{abstract}
\vspace{0.2cm}
\begin{IEEEkeywords}
Near-field channel estimation, polar-domain dictionary design, ultra-massive MIMO, sub-THz communications, hybrid architecture.\vspace{-0.2cm}
\end{IEEEkeywords}

\section{Introduction and Motivation}

\ac{U-MIMO} in the sub-THz band exploits the very short wavelengths to fit a large number of antenna elements into a compact area, which enables significant gains in beam focusing and spatial multiplexing \cite{bjornson2019massive}. This is envisioned as a key technology to meet the high traffic demands of wireless communications \cite{9166263,5764977,rappaport2019wireless}. However, in an all-digital architecture, where each antenna is paired with a dedicated \ac{RF} chain, this approach would be prohibitively power-intensive, demanding the use of hybrid architectures, where the number of \ac{RF} chains is limited \cite{9508929}.

In contrast to sub-$6$ GHz communications, where \acp{UE} are typically in the far field of the \ac{BS} array and orthogonal codebooks based on discrete Fourier transform matrices are effective, sub-THz frequencies place \acp{UE} in the radiating near field \cite{bacci2023spherical}. This shift requires the development of near-field codebooks, where the range domain becomes crucial due to the limited beamforming depth, introducing added complexity that goes beyond the angular considerations typical of far-field scenarios \cite{bjornson2021primer}. Accordingly, near-field codebooks are built sampling both the angular and range domains, resulting in a larger structure. Therefore, new channel estimation algorithms with overheads independent of dictionary size have emerged. Among them, the \ac{P-SOMP}\cite{cui2022channel} exploits channel sparsity for near-field channel estimation.

Existing codebooks fail to properly account for \ac{BS} height, as they are typically developed under the assumption that the \ac{BS} is positioned at ground level, which is almost never the case. In this paper, we propose a generalized near-field dictionary design criterion for arbitrary \ac{BS} location above the \ac{GP} where \acp{UE} are displaced. 

\begin{figure}[t]
    \centering
        \includegraphics[width =0.9 \columnwidth]{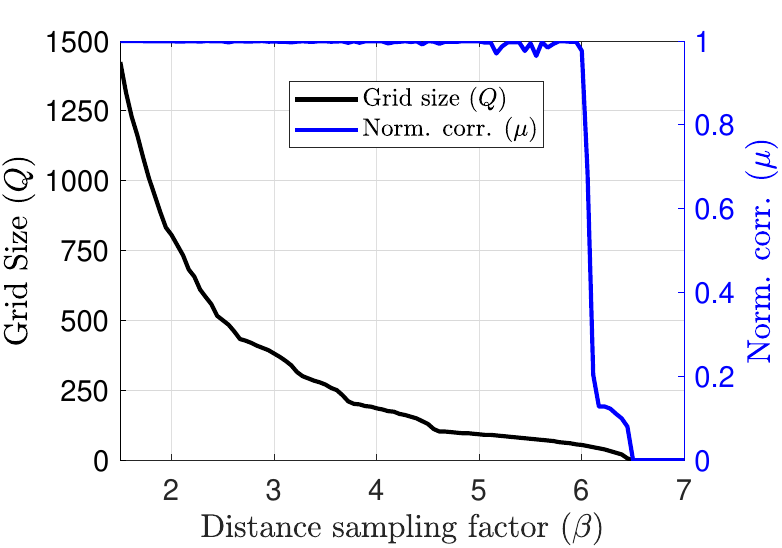}
        \caption{Normalized correlation $\mu$ and grid size $Q$ as a function of the distance sampling factor $\beta$ with the design in \cite{cui2022channel}.}\vspace{-0.5cm}
        \label{fig:1}
\end{figure}

 Additionally, recent designs of polar domain dictionaries for linear and planar arrays focus on the codebook's column coherence \cite{cui2022channel,demir2023new} - see Fig. \ref{fig:1}. However, since column coherence is not directly related to channel estimation, we use the \emph{optimal NMSE} to improve the codebook's accuracy. Results show that the proposed codebook outperforms state-of-the art criteria in terms of both channel estimation accuracy and \ac{SE} regardless of the codebook's size.

\section{System Model}
We consider the uplink of a U-MIMO system with $K$ single-antenna active \acp{UE}. We assume that the \ac{BS} is equipped with a \ac{ULA}. The number of antennas is $M$ and the inter-element spacing is $\delta$. The array is placed at a height $b$ above the \ac{GP}, where \acp{UE} are randomly displaced with uniform distribution within a given \ac{RoI}. We denote by $\mathbf{u}_m = \left[0, i(m)\delta,0\right]$ with $i(m) = \left(m - \frac{M-1}{2}\right)$ and $m = 0, 1 \ldots, M-1$ the coordinates of antenna $m$, and by $\mathbf{s}_k = (\rho_k\cos{\varphi_k},\rho_k\sin{\varphi_k},-b)$ with ${\rm \rho_{min}} \leq \rho_k \leq \rho_{\rm max}$ and ${\rm \varphi_{min}} \leq \varphi_k \leq \varphi_{\rm max}$ the coordinates of \ac{UE} $k$. Under \ac{LoS} propagation conditions, the electromagnetic channel from \ac{UE} $k$ to antenna $m$ is
\begin{align}
    h_{km} = \sqrt{\xi_{m}}e^{-j\frac{2\pi}{\lambda}r_{km}}
    \label{eq:SphericalModel} 
\end{align}
where ${\xi_{km}} \in \mathcal{CN}(0,\sigma_\xi^2)$ with 
\begin{align}
\sigma_\xi^2=\left(\frac{\lambda}{4\pi r_{km}}\right)^2
\end{align}
accounts for the path loss, $r_{km} = \| \mathbf{s}_k - \mathbf{u}_m \|$ denotes the Euclidean distance between \ac{UE} $k$ and antenna $m$ at the BS, and $\lambda$ is the wavelength.  We call $\mathbf{h}_k=\left[h_{k0},\dots,h_{k(M-1)}\right]^\mathsf{T}\in\mathbb{C}^M$ the channel vector of \ac{UE} $k$.
We consider a hybrid system in which the \ac{BS} has $K \leq N_{RF} \ll M$ \ac{RF} chains. Also, we assume a fully-connected architecture where each antenna is connected to all \ac{RF} chains \cite{8030501}.

\begin{figure}[t]
    \centering
    \begin{subfigure}{\columnwidth}
        \centering
        \includegraphics[width =0.9 \columnwidth]{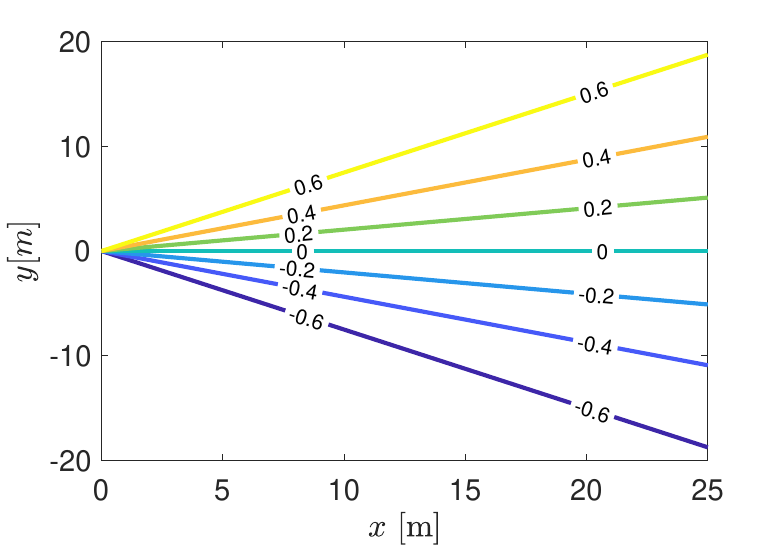}
        \caption{$b = 0$\,m.}
        \label{fig:2a}
    \end{subfigure}
    \begin{subfigure}{\columnwidth}
        \centering
        \includegraphics[width =0.9 \columnwidth]{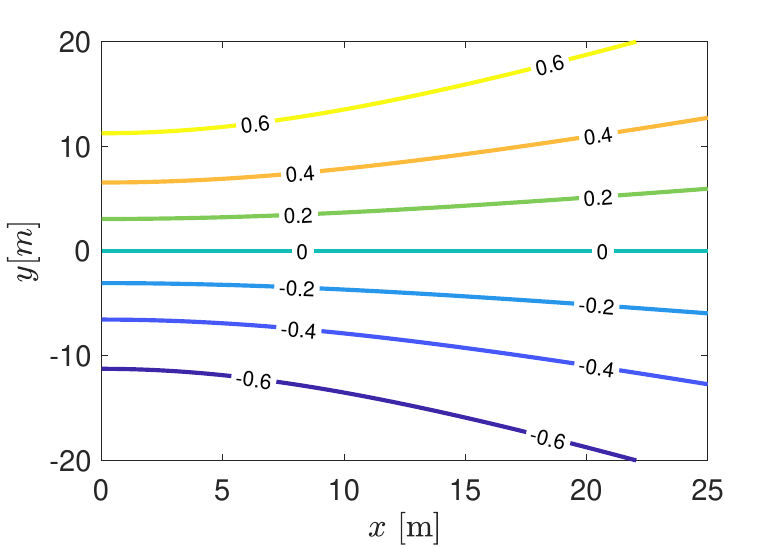}
        \caption{$b = 15$\,m.}
        \label{fig:2b}
    \end{subfigure}
       \caption{Level curves $\Gamma = g$ with seven different values of $g$ and \ac{BS} height $b = \{0,15\}$\,m.}
    \label{fig:2}
\end{figure}

\subsection{Channel Estimation}
The standard time-frequency division duplex protocol is used \cite{massivemimobook}, where $\tau_c$ channel uses are available for the uplink channel estimation phase and data transmission. We assume that the $K$ \acp{UE} simultaneously transmit over $K$ subcarriers, with separation achieved through orthogonal pilot sequences in the frequency-domain. The transmission occurs over $\tau$ pilot slots in the time domain. Hence, the number of resources used for channel estimation is equal to $\tau_p = K \tau$, and the remaining $\tau_c - \tau_p$ are used for data transmission. 

Signals at different antennas are combined by means of analogue combiners $\vect{a}_{i,r}\in \mathbb{C}^M$ for $i=1,\ldots,\tau$ and $r=1,\ldots,N_{RF}$, whose entries are randomly selected in $\{\pm 1/\sqrt{M}\}$ with equal probabilities, where the double-subscript notation indicates that different combiners in different time slots and different \ac{RF} chains may be used. After straightforward computations, the overall pilot signal during the channel estimation phase for each \ac{UE} can be written as
\begin{equation} \label{eq:pilot_vector}
    \vect{y} = \sqrt{p} K \vect{A} \vect{h} + \vect{n}
\end{equation}
where $p$ is the UE transmit power, $\vect{A} = [\mathbf{A}_1 \, \cdots \mathbf{A}_{\tau}]^{\Ttran}$ and $\vect{n} = [\vect{n}^{\Ttran}_1  \, \cdots \, \vect{n}^{\Ttran}_{\tau} ]^{\Ttran}$ where $\vect{A}_i = [\vect{a}_{i,1},\ldots, \vect{a}_{i,N_{RF}}] \in \mathbb{C} ^{M \times N_{RF}}$ is the analog combining matrix for the $i$th slot, and
$\vect{n}_{i} \in \mathcal{CN}(\vect{0},\sigma^2 K \vect{A}^{\mathsf{T}}_{i}\vect{A}_{i}^*)$ collects the noise.

We use the \ac{P-SOMP} algorithm, introduced in \cite{cui2022channel}, to estimate the channel by using the observation vectors in \eqref{eq:pilot_vector}. It is based on the polar-domain representation $\vect{h}^{\mathcal P}$ of the channel $\vect{h}$, i.e., $\vect{h} = \vect{W} \vect{h}^{\mathcal P}$ given in \cite[eq. (8)]{cui2022channel}, where $\vect{W} \in \mathbb{C}^{M \times Q} $ is a suitable \textit{dictionary}. More precisely, the $Q$ columns of $\vect{W}$ are the array steering vectors $\vect{w}(\theta,\varphi,\rho)$ \cite[eq. (7)]{cui2022channel} computed over a discrete grid $\vect{G}$ of {$(\theta,\varphi,\rho)$} values. Accordingly, they depend on the geometry of the array. A good design of $\vect{G}$ is essential for the operation of \ac{P-SOMP}.


\begin{figure}[t]
    \centering
        \includegraphics[width =0.9 \columnwidth]{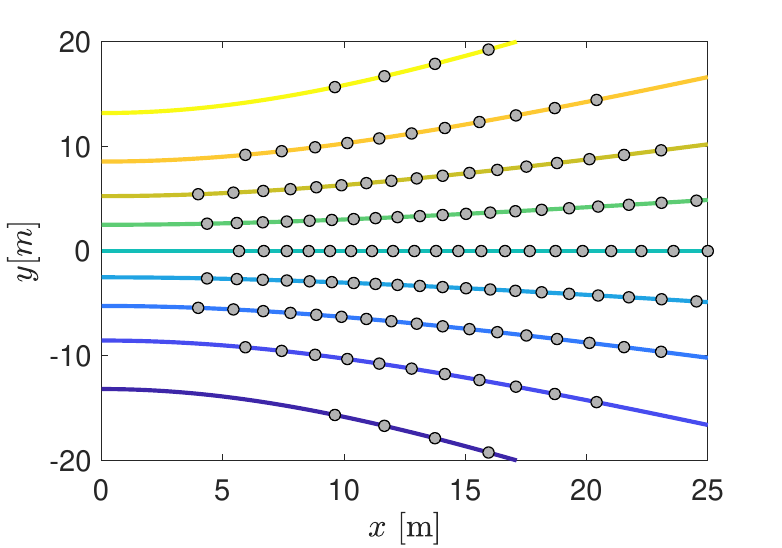}
        \caption{Example of the proposed grid construction with grid size $Q = 135$, number of level curves $N_\Gamma = 9$ and distance sampling factor $\beta = 0.57$.}
        \label{fig:3}
\end{figure}

\subsection{Review of the Dictionary Design from \cite{cui2022channel}} \label{sec:old_grid}
A possible grid design for a \ac{ULA} with $M$ antennas, lying on the ground plane $(b = 0)$, was presented in \cite{cui2022channel}. Particularly, \cite{cui2022channel} leads to a sampling of the angular domain with $M$ points such that $\Phi = \sin{\varphi}$ is computed over a uniform grid as
\begin{equation} \label{eq:cuidai_angular_sampling}
    \Phi = \frac{2m-M+1}{M} \quad m = 0,1,...,M-1.
\end{equation}
For each angle $\varphi$ compatible with \eqref{eq:cuidai_angular_sampling}, the distance is non-uniformly sampled as follows \cite[Eq. (15)]{cui2022channel}
\begin{equation} \label{eq:cuidai_distance_sampling}
    \rho_n = \frac{Z}{n} \quad n = 0,1,2 ...
\end{equation}
where
\begin{equation} \label{eq:cuidai_Z}
    Z = \frac{1}{2\lambda}\left(\frac{M \delta}{\beta}\right)^2\left(1-\Phi^2\right)
\end{equation}
and $\beta$ is a design parameter \cite{cui2022channel}. In \cite{cui2022channel}, $\beta$ is designed 
by considering the maximum correlation between columns of the dictionary $\vect{W}$, defined as
\begin{equation} \label{eq:max_corr}
    \mu = \underset{i \neq j}{\text{max}}\left\{|\vect{w}_i^H\vect{w}_j|\right\}
\end{equation}
where $\vect{w}_i$ and $\vect{w}_j$ are two columns of $\vect{W}$. On one side, the maximum correlation should be made as small as possible for limiting the number of grid points. On the other side, it cannot be too small because this would lead to a very sparse grid $\vect{G}$, with very few number of points and a reduced estimation accuracy. Fig.~\ref{fig:1} shows $\mu$ (blue line), normalized by the number of antennas $M$, and the dictionary size $Q$ (black line) as a function of $\beta$ for a \ac{ULA}. The blue curve is obtained using the exact (spherical) electromagnetic model for the array steering vectors. We see that increasing $\beta$ results in smaller values of $\mu$ and in a smaller grid size. Notably, the only grid design parameter in \cite{cui2022channel} is the distance sampling factor $\beta$ in \eqref{eq:cuidai_distance_sampling}. Hence, we can choose $\beta$ to either achieve a given grid size $Q$ or a given normalized correlation $\mu$ in \eqref{eq:max_corr}, but not both. 


\begin{figure}[t]
    \centering
    \includegraphics[width = .9\columnwidth]{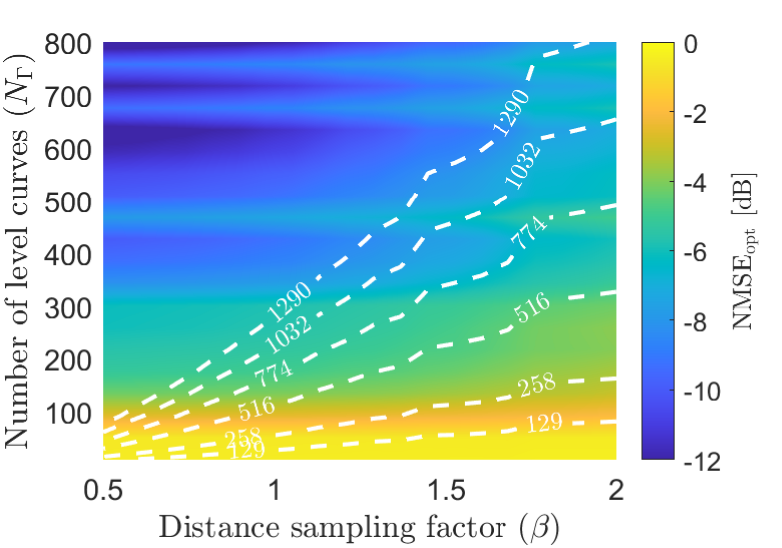}
    \caption{Proposed design with \ac{BS} height $b = 15\,$m.}
    \label{fig:4}
\end{figure}

\section{Proposed polar dictionary design} \label{sec:proposed_design}
Our design of the grid is based on the same line of reasoning as in \cite{cui2022channel}, but we also take into account the elevation of the array with respect to the \ac{GP}. Details are reported in the Appendix, where the concept of a \textit{level curve} is introduced. The latter is defined as the set of points in the \ac{GP} for which
\begin{equation} \label{eq: LevelCurve_simply}
\Gamma(R,\varphi) \overset{\rm def}{=} \cos{\theta}\sin{\varphi} = \sqrt{1 - \left(\dfrac{b}{R}\right)^2} \sin{\varphi} = g
\end{equation}
where $g \in [-1,1]$ is an arbitrary constant, and $R$ denotes the distance from the center of the \ac{BS} array. Fig.~\ref{fig:2} shows the level curves in \eqref{eq: LevelCurve_simply} for $g = 0, \pm 0.2, \pm 0.4, \pm 0.6,$ and $b = 0,15$\,m. When $b = 0$, the level curves appear as straight lines; as the BS elevation $b$ increases, these curves progressively bend, reflecting the growing influence of elevation. 

\subsection{Proposed level curves}

Level curves are selected such that $\Gamma(R,\varphi) = \Gamma_k$ with 
\begin{equation} \label{eq:Gamma_k_A}
\Gamma_k = \Gamma_{\rm max}\left(\frac{2k-N_\Gamma+1}{N_\Gamma}\right) \quad k=0,1,\ldots, N_\Gamma-1
\end{equation}
where 
\begin{equation}
\Gamma_{\max} = \frac{\rho_{\max}}{ \sqrt{\rho^2_{\max} + b^2}} \sin \varphi_{\max} 
\end{equation}
and $\{\varphi_{\max},\rho_{\max}\}$ denote the maximum azimuth and radial distance within the \ac{RoI}, respectively.
The condition $\rm |\Gamma| < \Gamma_{max}$ ensures that all used level curves pass through the \ac{RoI}.

Note that the level curves in \eqref{eq:Gamma_k_A} play the same role as the straight lines in \eqref{eq:cuidai_angular_sampling}. They coincide when the elevation is zero $(b = 0)$, $\rm \varphi_{max} = \frac{\pi}{2}$ and $N_\Gamma = M$. Unlike in \eqref{eq:cuidai_angular_sampling}, where the number of level curves $N_\Gamma$ is fixed to $M$, we consider $N_\Gamma \geq M$, thereby introducing $N_\Gamma$ as an additional design parameter. When $N_\Gamma > M$, more samples than in \cite{cui2022channel} are considered in the angular domain in spite of the distance domain. We call this \emph{angular domain over-sampling} and discuss its advantages on channel estimation accuracy.

\begin{table}[t]
    \renewcommand{\arraystretch}{1.2}
    \centering
    \begin{tabular}{c|c}
    \textbf{Parameter} & \textbf{Value} \\
    \hline
    Carrier Frequency, $f_c$ & $300\,\, [\mathrm{GHz}]$ \\ \hline
    Bandwidth, $B$ & $100\,[\mathrm{MHz}]$ \\ \hline 
    Array Aperture, $L$ & $0.64$ [m]\\ \hline
    Number of Antennas, $M$ & $129$ \\ \hline
    Antenna Spacing, $\delta$ & $5\lambda$ \\ \hline
     Minimum UE distance, $\rho_{\min}$ & $5$ [m] \\ \hline
     Maximum UE distance, $\rho_{\max}$ & $25$ [m] \\ \hline
    Minimum UE azimuth $\varphi_{\min}$ & $-\pi/3$ [rad] \\ \hline
    Maximum UE azimuth $\varphi_{\max}$ & $\pi/3$ [rad] \\ \hline
    \ac{BS} height, $b$ & $0,15$ [m] \\
    \end{tabular}
    \caption{System parameters.}
    \label{tab:system_parameters}
\end{table}

\begin{table}[t]
    \renewcommand{\arraystretch}{1.2}
    \centering
    \begin{tabular}{c|c}
        \textbf{Parameter} & \textbf{Value} \\ \hline
        Pilot Sequence Length, $\tau_p$ & $10$ \\ \hline
        Number of RF chains, $N_{RF}$ & $10$ \\ \hline
        Transmit Power, $p$ & $15$ [dBm] \\ \hline
        Thermal Noise Power, $\sigma^2$ & $-86$ [dBm] \\ \hline
        Number of UEs, $K$ & $10$ \\ \hline
        UEs distribution & Uniform in the \ac{RoI} \\  \hline
        Dictionary size, $Q$ & $4M$,$10M$ \\
    \end{tabular}
\caption{Channel estimation parameters.}
\label{tab:channel_estimation_parameters}
\end{table}

\subsection{Proposed distance sampling}

To determine the distance sampling, we select the grid points on each level curve within the \ac{RoI}. In particular, those belonging to the level curve $\Gamma(R, \varphi) = \Gamma_k$ are selected so that
\begin{equation} \label{eq:DistanceSampling}
    R_{n,k} = \dfrac{Z_k R_{0}}{Z_k + nR_{0}} \quad \quad n=0,1,2,\ldots
\end{equation}
where 
\begin{equation} \label{eq:Z_k}
Z_{k} = \frac{1}{2\lambda}\left(\frac{M\delta}{\beta}\right)^2\left(1-\Gamma_k^2\right)
\end{equation}
$\beta$ is a distance sampling factor, and $R_0$ is the arbitrary maximum distance considered on each level curve that we set as $R_0 = \sqrt{ \rho_{\rm max}^2 + b^2}$. Notice that distances $R_{n,k}$ in \eqref{eq:DistanceSampling} play the same role as radial distances $\rho_s$ in \eqref{eq:cuidai_distance_sampling}. They coincide when the elevation is zero $(b = 0)$, $\rm \varphi_{max} = \frac{\pi}{2}$, $N_\Gamma = M$ and $R_0 \to \infty$.

\subsection{Grid construction}
The proposed grid construction can be summarized as follows:
\begin{enumerate}[label = \arabic*.]
    \item Set the grid size $Q$.
    \item Determine the number of level curves $N_\Gamma$ and the distance sampling factor $\beta$.
    \item Compute the level curves $\Gamma(R, \varphi) = \Gamma_k$ from \eqref{eq:Gamma_k_A}.
    \item Compute the distances $R_{n,k}$ using \eqref{eq:DistanceSampling}.
    \item Compute grid points coordinates on each level curve as
    \begin{equation} \label{eq:grid_point_coordinates}
        \left[\sqrt{R_{n,k}^2(1-\Gamma_k^2)-b^2},R_{n,k}\Gamma_k,-b\right]^T.
    \end{equation}
\end{enumerate}

Fig.~\ref{fig:3} shows an example of the proposed grid construction with $Q = 135$, $N_\Gamma = 9$ and $\beta = 0.57$.

\begin{figure}[t] \vspace{-0.5cm}
    \centering
    \begin{subfigure}{\columnwidth}
        \centering
        \includegraphics[width = 0.9 \columnwidth]{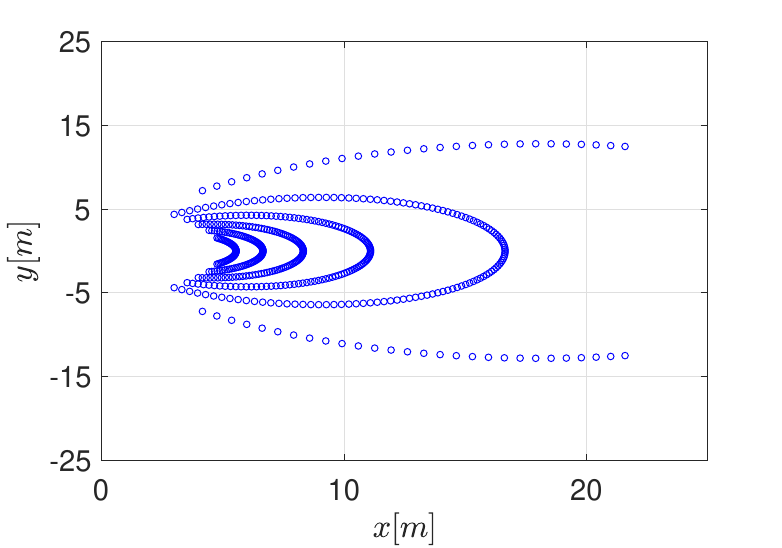}
        \caption{$\beta = 2.5$, $Q = 501$.}
        \label{fig:4a}
    \end{subfigure}
    \begin{subfigure}{\columnwidth}
        \centering
        \includegraphics[width =0.9 \columnwidth]{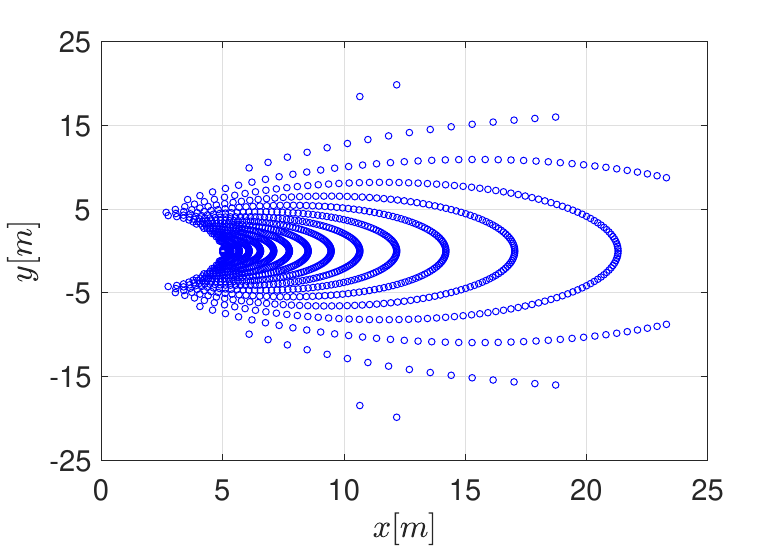}
        \caption{$\beta = 1.56$, $Q = 1298$.}
        \label{fig:5b}
    \end{subfigure}
    \caption{Polar-domain grids obtained using the design in \cite{cui2022channel} with $\beta = \{2.5,1.56\}$ and $Q = \{501,1298\}$, respectively.}
    \label{fig:5}
\end{figure}

    \begin{figure}[t] \vspace{-0.5cm}
    \centering
    \begin{subfigure}{\columnwidth}
        \centering
        \includegraphics[width = 0.9 \columnwidth]{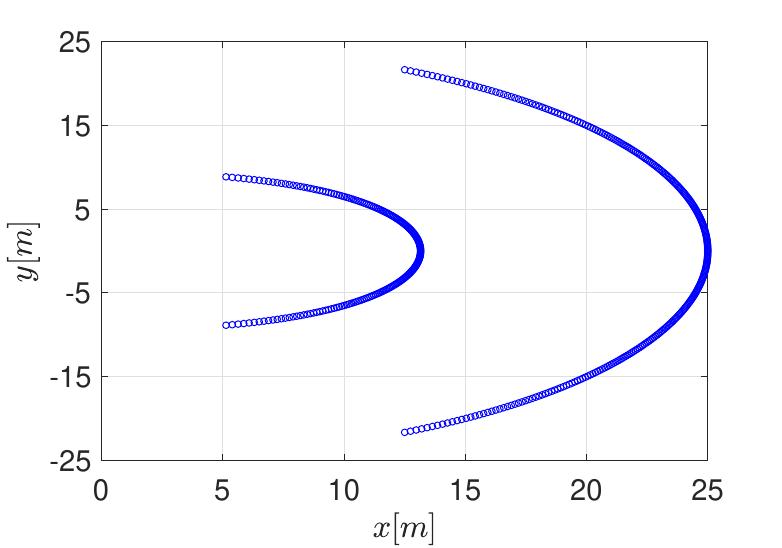}
        \caption{$N_\Gamma = 310$, $\beta = 1.81$, $Q = 514$.}
        \label{fig:6a}
    \end{subfigure}
    \begin{subfigure}{\columnwidth}
        \centering
        \includegraphics[width =0.9 \columnwidth]{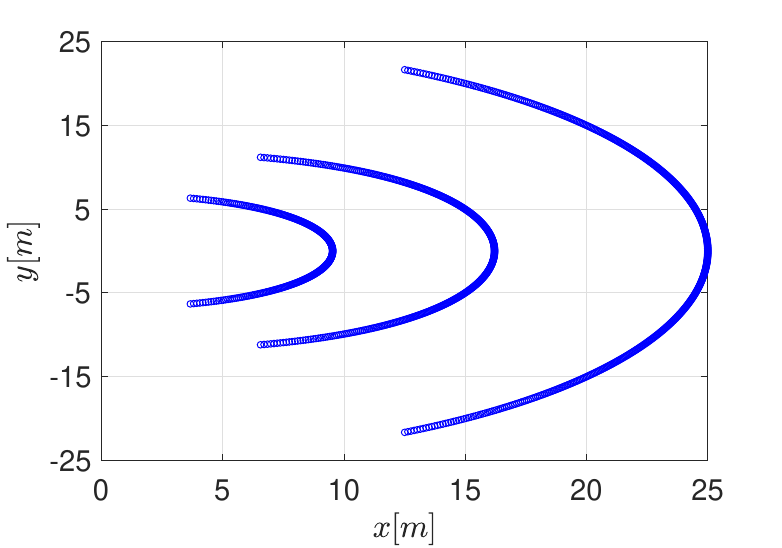}
        \caption{$N_\Gamma = 568$, $\beta = 1.51$, $Q = 1298$.}
        \label{fig:6b}
    \end{subfigure}
    \caption{Polar-domain grids obtained using the proposed design with $N_\Gamma = \{310,568\}$, $\beta = \{1.81,1.51\}$ and $Q = \{514,1290\}$, respectively.}
    \label{fig:6}
\end{figure}

\subsection{Grid design}
The column coherence in \eqref{eq:max_corr} is not directly related to channel estimation. Hence, unlike \cite{cui2022channel}, we design $N_\Gamma$ and $\beta$ according to the optimal NMSE, defined as  
\begin{equation} \label{eq:NMSE_opt}
    \text{NMSE}_\text{opt} = 1 - {\mathbb E}\left\{\underset{\vect{g} \in \vect{G}}{\text{min}}\left[\frac{|\vect{h}^H(\vect{g})\vect{h}(\vect{r})|^2}{||\vect{h}(\vect{g})||^2||\vect{h}(\vect{r})||^2}\right]\right\}
\end{equation}
where $\vect{h}({\vect r})$ indicates the channel of the UE located in $\bf r$, and $\vect{h}({\vect g})$ is the channel from the point $\vect{g} \in \vect{G}$. The expectation is computed with respect to the {UE}'s position ${\vect r}$. Note that $\text{NMSE}_\text{opt}$ can be interpreted as a measure of the average error in approximating the channel at an arbitrary point with the channel at a nearby grid point. From this perspective, lower values of $\text{NMSE}_\text{opt}$ are expected to improve the estimation accuracy for grid-based algorithms, such as the \ac{P-SOMP}.  
To understand how to choose $N_\Gamma$ and $\beta$, let us consider Fig.~\ref{fig:4}, which shows the $\text{NMSE}_\text{opt}$ in \eqref{eq:NMSE_opt} as a function of both $\beta$ and $N_{\Gamma}$. All other simulation parameters are summarized in Tabs.~\ref{tab:system_parameters} and~\ref{tab:channel_estimation_parameters}. It can be observed that, for any given grid size $q$, there exist multiple pairs $(N_\Gamma, \beta)$ lying on the level curve corresponding to the constant grid size $Q = q$ (indicated by white dashed lines), each yielding the desired grid dimension $q$. Among these, we choose the pair $(N_\Gamma,\beta)$ that minimizes the $\rm NMSE_{opt}$ in \eqref{eq:NMSE_opt}.

\section{Performance evaluation}
We now evaluate the \ac{P-SOMP} algorithm in terms of both channel estimation accuracy and \ac{SE}. The communication happens at the carrier frequency $f_c = 300$\,GHz, which yields $\lambda = 1$\,mm, over a bandwidth $B = 100$\,MHz. The $M = 129$ antennas at the \ac{BS} are spaced by $\delta = 5\lambda = 5$\,mm, which yields an array aperture $L \approx 0.64$\,m. The $K = 10$ \acp{UE} are randomly displaced with uniform distribution within the \ac{RoI}, which is shaped as the circular sector on the \ac{GP} with minimum and maximum azimuth angles and radial distances given by $\rm \varphi_{max} = -\varphi_{min} = \frac{\pi}{2}$, $\rm \rho_{max} = 25$\,m and $\rm \rho_{min} = 5$\,m, respectively. The \ac{BS} is located at height $b = 15$\,m above the \ac{GP}. The transmit power of each user is $p \in [0,20]$\,dBm, and the thermal noise power $\sigma^2 = -86$\,dBm. Eventually, we consider both a smaller and a larger grid size, which are $Q = \{4M,10M\}$, respectively.
With these two values of $Q$, the design in \cite{cui2022channel} yields $\beta = \{2.5,1.56\}$, while the proposed method results in $N_\Gamma = \{310,568\}$ and $\beta = \{1.8,1.51\}$, respectively. Figs.~\ref{fig:5} and~\ref{fig:6} show the corresponding grids. Notably, despite the grids sharing approximately the same size, the number of level curves is fixed to $N_\Gamma = M$ in \cite{cui2022channel}, while it is $N_\Gamma > M$ in both proposed grids, resulting in a clear angular domain over-sampling. All simulation parameters are summarized in Tables~\ref{tab:system_parameters} and~\ref{tab:channel_estimation_parameters}.

\begin{figure}[t]
    \centering
        \includegraphics[width = 0.9 \columnwidth]{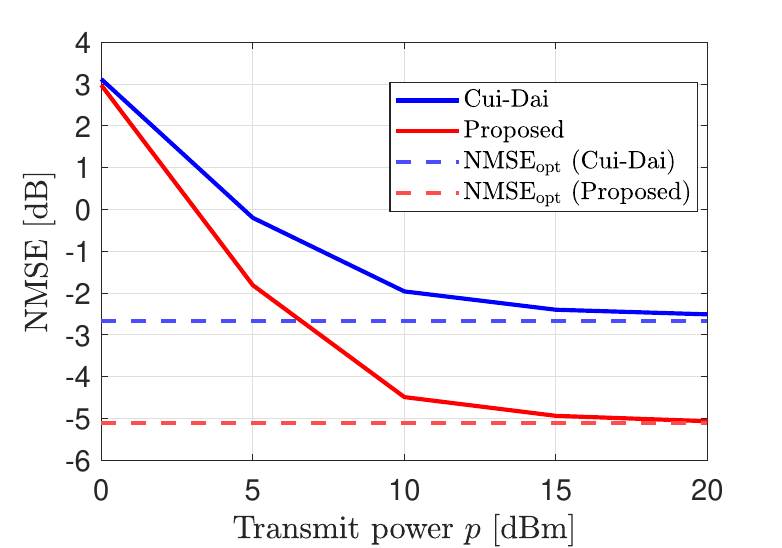}
        \caption{NMSE and $\rm NMSE_{opt}$ as functions of the transmit power $p$ in dBm, with grid size $Q = 4M$.}
        \label{fig:7}
\end{figure}

We evaluate the channel estimation accuracy in terms of both $\rm NMSE_{opt}$ in \eqref{eq:NMSE_opt} and actual NMSE, achieved with the \ac{P-SOMP} algorithm. Firstly, we consider $Q = 4M$ and report results as a function of the transmit power $p$. Fig.~\ref{fig:7} shows that both $\rm NMSE_{opt}$ and NMSE improve up to $3$\,dB using the proposed design instead of the one in \cite{cui2022channel}. Additionally, this shows that angular domain oversampling can lead to more precise channel estimation.

The \ac{P-SOMP} accuracy improves with larger dictionaries. For example, by considering a grid size $Q = 10M$, both the dictionary design in \cite{cui2022channel} and the proposed one yield more accurate channel estimations. Specifically, $\rm NMSE_{opt}$ decreases to $-5$\,dB and $-8.7$\,dB, while the actual NMSE improves to $-4.2$\,dB and $-7.7$\,dB, respectively. However, this comes at the price of higher computational complexity of the \ac{P-SOMP}, which grows linearly with $Q$ \cite{cui2022channel}. 


\begin{figure}[t]
    \centering
        \includegraphics[width = 0.9 \columnwidth]{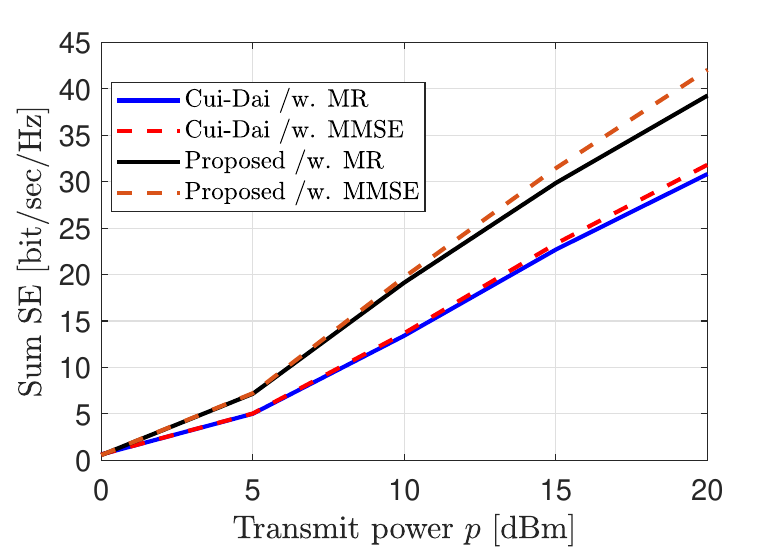}
        \caption{Sum SE as a function of the power $p$ with $K = 10$ \acp{UE} and grid size $Q = 4M$. Both MR and MMSE combiners are considered.}
        \label{fig:8}
\end{figure}


We now evaluate the achievable sum uplink \ac{SE} with both \ac{MR} and \ac{MMSE} combining schemes. This is obtained with the well-known use-and-then-forget bound \cite[Sec.~4.2]{massivemimobook}. Fig.~\ref{fig:8} reports the sum SE as a function of $p$ with $Q = 4M$. We see that \ac{SE} is higher with the proposed design across the whole considered power range. For example, \ac{SE} approximately improves by $6$\,bit/sec/Hz for $p \geq 10$\,dBm. The gain on \ac{SE} increases up to $8$\,bit/sec/Hz with $Q = 10M$.

\section{Conclusions}
We introduced a novel dictionary design that extends the traditional correlation-based criterion—commonly applied under far-field assumptions—by explicitly accounting for the BS height. Unlike existing approaches, which typically assume the BS is located at ground level, our design generalizes the grid construction to accommodate elevated BS deployments. Furthermore, the grid is dimensioned based on the minimization of the optimal NMSE, rather than the conventional correlation-based metric, enabling a more accurate representation of channel characteristics. The proposed grid is integrated into a hybrid U-MIMO system operating at $300$ GHz, employing the P-SOMP algorithm~\cite{cui2022channel} for channel estimation. Simulation results demonstrate that our design consistently outperforms existing grid strategies in terms of both channel estimation accuracy and SE.

\section*{Appendix}
We begin with the spherical coordinates of the generic grid point $\mathbf{p}$, given by
\begin{equation} \label{eq:grid_point_spherical}
    \mathbf{p} =[R\cos{\theta}\cos{\varphi},R\cos{\theta}\sin{\varphi},-b]^T 
\end{equation}
where 
\begin{equation} \label{eq:cos_theta}
    \cos{\theta} = \sqrt{1 - \left(\frac{b}{R}\right)^2}.
\end{equation}
The second-order (i.e., parabolic) approximation of the distance $r_{m} = \| \mathbf{p} - \mathbf{u}_m \|$ is given by:
\begin{align} \label{eq:ParabolicAppx}
r_m \approx &- \delta \left(\cos \theta \sin\varphi \right) i(m) \nonumber \\ &+ \dfrac{\delta^2}{2R} \left( 1 - \cos^2 \theta  \sin^2 \varphi \right) i^2(m).
\end{align}
The associated steering vector can be written as
\begin{equation}
\label{eq: SteeringVector}
\vect{s}(R,\varphi,b) = \left[e^{\imagunit\frac{2 \pi}{\lambda}\|\vect{p} - \vect{u}_0\|},\ldots,e^{\imagunit\frac{2 \pi}{\lambda}\|\vect{p} - \vect{u}_{M-1}\|}\right]^\Ttran.
\end{equation}
Now, let us consider two grid points $\mathbf{p}$ and $\mathbf{q}$ with coordinates
\begin{align}
    \mathbf{p} &= [R_p\cos{\theta_p}\cos{\varphi_p},R_p\cos{\theta_p}\sin{\varphi_p},-b]^T, \\
    \mathbf{q} &= [R_q\cos{\theta_q}\cos{\varphi_q},R_q\cos{\theta_q}\sin{\varphi_q},-b]^T.
\end{align} Based on \eqref{eq:ParabolicAppx}, the modulus of the correlation between the corresponding steering vectors (after straightforward computations) can be approximated as \cite[eq. (9)]{cui2022channel}:
\begin{equation}
\label{eq: Correlation1}
f(\varphi_p, \varphi_q, R_{p}, R_{q}) \approx \left | \dfrac{1}{M} \sum_{m = - (M-1)/2}^{(M-1)/2} e^{\iu (A m + B m^2)}\right |
\end{equation}
where
\begin{equation} \label{eq: A}
A = \dfrac{2 \pi \delta} {\lambda} (\Gamma_q -\Gamma_p)
\end{equation}
\begin{equation} \label{eq: B}
B = \dfrac{\pi \delta^2} {\lambda} \left[ \dfrac{1} {R_p}\left(1 - \Gamma_p^2 \right) - \dfrac{1} {R_q}\left(1 - \Gamma_q^2 \right) \right]
\end{equation}
with 
\begin{equation} \label{eq:Gamma}
\Gamma_i = \cos{\theta_i}\sin \varphi_i = \sqrt{1-\left(\frac{b}{R_i}\right)^2} \sin \varphi_i    \quad \quad i = p,q   
\end{equation}

Now, assume that the two points $\mathbf{p}$ and  $\mathbf{q}$ lie on the same level curve, i.e., $\Gamma_p = \Gamma_q = g$. In this case, equations~\eqref{eq: A} and~\eqref{eq: B} reduce to
\begin{equation}
\label{}
A = 0  \quad \quad B = \dfrac{\pi \delta^2}{\lambda}(1 - g^2) \left(\dfrac{1}{R_p} - \dfrac{1}{R_q}\right) 
\end{equation}
Accordingly, the correlation $f(\varphi_p, \varphi_q, R_{p}, R_{q})$ in \eqref{eq: Correlation1} becomes
\begin{equation}
\label{eq: Correlation2}
f(\varphi_p, \varphi_q, R_{p}, R_{q}) = \left | \dfrac{1}{M} \sum_{m = -(M-1)/2}^{(M-1)/2} e^{\iu B m^2}\right |
\end{equation}
which can be further approximated as in \cite[Eq. (12)]{cui2022channel}:
\begin{equation}
\label{eq: Correlation3}
f(\varphi_p, \varphi_q, R_{p}, R_{q}) \approx \left| G(\beta) \right| = \left | \dfrac{C(\beta) + \iu S(\beta)}{\beta} \right |
\end{equation}
where $C(\beta)$ and $S(\beta)$ are the cosine and sine Fresnel integral functions \cite{1137900}, respectively, and
\begin{equation}
\label{eq: beta}
\beta^2 = \dfrac{M^2 \delta^2 (1 - g^2)}{2 \lambda} \left|\dfrac{1}{R_p} - \dfrac{1}{R_q}\right|.
\end{equation}
Equation \eqref{eq: Correlation3} shows that, by varying $\beta$, we may control the correlation between the channels (actually, the steering vectors) in $\mathbf{p}$ and $\mathbf{q}$. Additionally, equation~\eqref{eq: beta} leads to the distance sampling criterion in \eqref{eq:DistanceSampling} and \eqref{eq:Z_k}, where $R_0$ is an additional design parameter denoting the arbitrary maximum distance from grid points to the \ac{BS} on each level curve.
Unlike \cite{cui2022channel}, where $R_0 \to \infty$, we consider $R_0 = \sqrt{\rm \rho_{max}^2 + b^2}$, taking into account that our \ac{RoI} is limited.

In order to construct the grid, we need to fix the number of level curves $N_\Gamma$ and their values. To this end, observe that the difference $\Gamma_p - \Gamma_q$ determines the correlation  $f(\varphi_p, \varphi_q, R_{p}, R_{q})$ in the far field, when $R_{p} \to \infty$ and  $R_{q} \to \infty$, as can easily be derived from \eqref{eq: Correlation1}-\eqref{eq: B}. Indeed, in this case $B \to 0$ and \eqref{eq: Correlation1} reduces to
\begin{equation} \label{eq: Correlation4}
f(\varphi_p, \varphi_q, R_{p}, R_{q}) \approx \left| \dfrac{\sin\left[\dfrac{\pi M \delta}{\lambda}(\Gamma_p - \Gamma_q)\right]} {M \sin\left[\dfrac{\pi \delta}{\lambda}(\Gamma_p - \Gamma_q)\right]}\right|
\end{equation}
This suggests considering values of $\Gamma$ uniformly spaced in the interval $( - \Gamma_{\max}, \Gamma_{\max})$, which leads to the level curves selection criterion in \eqref{eq:Gamma_k_A}.

Once the level curves $\Gamma(R,\varphi) = \Gamma_k$ and the distances $R_{n,k}$ on each level curve have been chosen according to \eqref{eq:Gamma_k_A} and \eqref{eq:DistanceSampling}, respectively, we compute the spherical coordinates of each grid point in \eqref{eq:grid_point_spherical} as per \eqref{eq:grid_point_coordinates}.

\section*{Acknowledgment}
This work has been performed in the framework of the HORIZON-JU-SNS-2022 project TIMES, cofunded by the European Union. Views and opinions expressed are however those of the author(s) only and do not necessarily reflect those of the European Union.

\bibliographystyle{IEEEtran}
\bibliography{IEEEabrv,refs}

\end{document}